\documentclass{ws-procs975x65}
\usepackage{amsmath, amssymb}

\def\Mpch{~h^{-1} {\rm Mpc}}

\newcommand{\nexus}{NEXUS+~}
\newcommand{\nexuS}{NEXUS+}
\newcommand{\logFilter}{Log-Gaussian~}
\newcommand{\lcdm}{$\Lambda$CDM~}

\newcommand{\reffig}[1]{Fig. \ref{#1}}

\begin{document}

\title{\uppercase{Nexus of the Cosmic Web}}

\author{\uppercase{Marius Cautun}$^*$, \uppercase{Rien van de Weygaert} and \uppercase{Bernard J.T. Jones}}
\address{Kapteyn Astronomical Institute, University of Groningen,\\
Groningen, 9747 AD, The Netherlands\\
$^*$E-mail: cautun@astro.rug.nl}

\author{\uppercase{Carlos S. Frenk} and \uppercase{WOJCIECH A. HELLWING} }
\address{Department of Physics, Institute for Computational Cosmology, University of Durham, \\
South Road Durham DH1 3LE, U.K.}

\begin{abstract}
One of the important unknowns of current cosmology concerns the effects of the large scale distribution of matter on the formation and evolution of dark matter haloes and galaxies. One main difficulty in answering this question lies in the absence of a robust and natural way of identifying the large scale environments and their characteristics. This work summarizes the NEXUS+ formalism which extends and improves our multiscale scale-space MMF method. The new algorithm is very successful in tracing the Cosmic Web components, mainly due to its novel filtering of the density in logarithmic space. The method, due to its multiscale and hierarchical character, has the advantage of detecting all the cosmic structures, either prominent or tenuous, without preference for a certain size or shape. The resulting filamentary and wall networks can easily be characterized by their direction, thickness, mass density and density profile. These additional environmental properties allows to us to investigate not only the effect of environment on haloes, but also how it correlates with the environment characteristics.
\end{abstract}


\bodymatter

\bigskip

Galaxy redshift surveys reveal that the distribution of galaxies in the universe is not random\cite{Gregory78,Geller89}, but forms a large scale pattern referred to as the \textit{Cosmic Web}\cite{Bondweb96}. Recent studies show that there is a correlation between the large scale environment and the properties of dark matter haloes in N-body simulations\cite{Aragon07a,Hahn2007a} and that of galaxies in redshift surveys\cite{Jones10,2012arXiv1207.0068T}. To get a better understanding of what are the causes of these effects and how they arise, we need a method that identifies in a natural and robust way the components of the Cosmic Web and their features. Tracing the Cosmic Web is challenging due to the characteristics of the megaparsec matter distribution: hierarchical composition with both large and small structures, no clear boundaries between the different components and a large range in density between the underdense and overdense regions. While there already exists a broad set of methods for identifying the Cosmic Web, most of these tools fail in capturing at least one of the aspects outlined above.

In Cautun et al. (2012)\cite{cautun2012} we present the \nexus algorithm for the segmentation of the Cosmic Web into its distinct morphological components: clusters, filaments, walls and voids. The \nexus formalism is an extension and improvement of our multiscale scale-space formalism, which we introduced in Aragon-Calvo et al. (2007)\cite{Aragon07b}. The method properly takes into account all the aspects of megaparsec matter distribution described above by following a multiscale and parameter free approach. \nexus performs the Cosmic Web identification using the Hessian eigenvalues of the smoothed density field. The \nexus algorithm can be summarized in the following six steps:

\begin{figure}
    \begin{center}
        \psfig{file=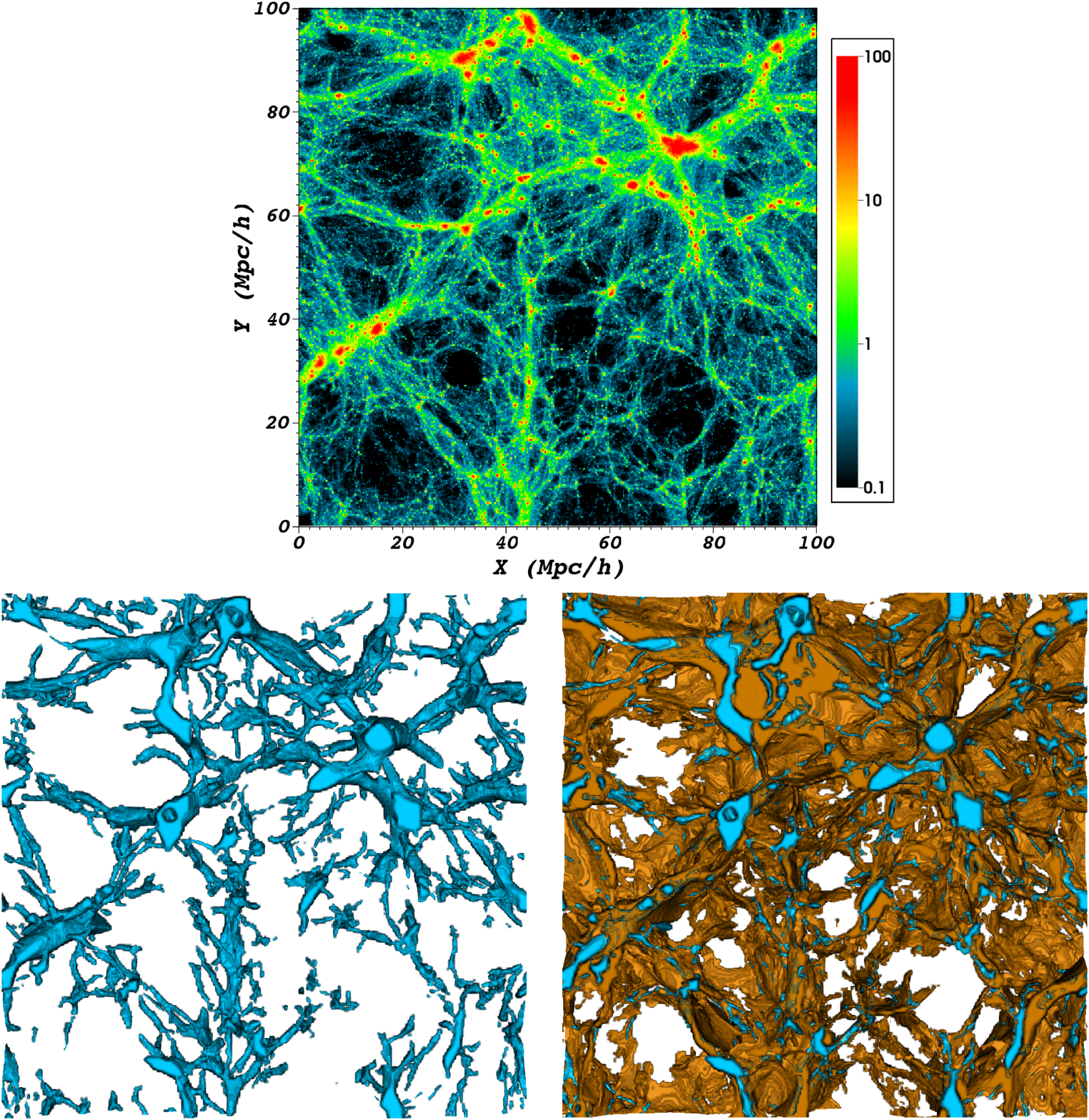,angle=0,width=.8\textwidth}
        \caption{Upper panel: A $10\Mpch$ density slice through the results of a \lcdm N-body simulation at present redshift. The legend on the right gives the color-code corresponding to the overdensity $1+\delta$. Lower panel: A 3D rendering of the \nexus filament and wall environments detected in the volume of the density slice. The left panel shows only the filaments in blue while in the right panel on top of the filaments we superimpose the walls in orange.}
        \label{fig:env}
    \end{center}
\end{figure}

\begin{itemize}
    \item[(I)] Applying the \logFilter filter of width $R_n$ to the input density field. The \logFilter smoothing consists of a Gaussian filter that acts on the logarithm of the density.
    \item[(II)] Computing the Hessian matrix eigenvalues for the filtered density field.
    \item[(III)] Assigning a cluster, filament and wall signature to each point using the Hessian eigenvalues determined in the previous step. The environment signature is defined on the basis of the morphological characteristics associated to clusters, filaments and walls.
    \item[(IV)] Repeating steps \textit{(I)} to \textit{(III)} over a set of smoothing scales $(R_0,R_1,..,R_N)$. The result of this step is a set of environmental signatures for each scale.
    \item[(V)] Combining the results of all scales to obtain a scale independent cluster, filament and wall signature. 
    \item[(VI)] Using physical criteria to determine the detection threshold corresponding to valid environments.
\end{itemize}

To exemplify the strength of the method we applied it to the detection of the Cosmic Web in a \lcdm N-body simulation. As input for the \nexus method we used the DTFE\cite{2000A&A...363L..29S,Cautun2011} interpolated density field. The resulting filaments and walls through a $10\Mpch$ slice are shown in the lower panel of \reffig{fig:env}. For comparison we plot the density field in the upper panel of the same figure. The region in \reffig{fig:env} was selected to included the most massive halo in the simulation to better illustrate the complexity of the filamentary and sheet networks that are found by \nexuS. We find that the environments identified by \nexus have a one to one correspondence to the structures that are visible in the density field and the method recovers the natural size, extent and shape of the large scale structures. We see that the cosmos is criss-crossed with thick and massive filaments which branch into thinner and thinner filaments that pierce the cosmic voids. The filaments are embedded in sheets, with the thick ones lying at the intersection of prominent walls while the thin ones are located in tenuous walls.

We find that due to its scale-space approach and use of the \logFilter filter, this method is equally sensitive in the detection of both prominent and tenuous filaments and walls. Moreover \nexus identifies the natural shape and extent of the structures which makes it an ideal tool for measuring the size, density and density profile of the Cosmic Web components. This allow us not only to better understand the formation and evolution of the Cosmic Web, but also to measure its characteristics. This is especially useful when making an extensive analysis of the halo properties dependence on environment and especially on environment characteristics.

\newcommand{\araa}{Annual Review of Astronomy and Astrophysics}
\newcommand{\apj}{APJ}
\newcommand{\nat}{Nature}
\newcommand{\apjs}{APJS}
\newcommand{\apjl}{APJL}
\newcommand{\mnras}{MNRAS}
\newcommand{\aap}{A\&A}

\bibliographystyle{ws-procs975x65}
\bibliography{main}
\end{document}